\documentclass[12pt, aaspp4, flushrt, mathptmx]{aastex}

\usepackage{epsf}
\usepackage{verbatim}

\newcommand{\ha}{H$\alpha$}
\newcommand{\comments}[1]{}

\begin{document}

\sffamily

\title{Dynamics of Chromospheric Upflows and Underlying Magnetic Fields}

\author{Yurchyshyn, V., Abramenko, V., Goode, P.}

\affil{\it Big Bear Solar Observatory, New Jersey Institute of Technology, Big Bear City, CA 92314, USA}

\begin{abstract}
We used \ha-0.1~nm and magnetic field (at 1.56$\mu$) data obtained with the New Solar Telescope to study the origin of the disk counterparts to type II spicules, so-called rapid blueshifted excursions (RBEs). The high time cadence of our chromospheric (10~s) and magnetic field (45~s) data allowed us to generate x-t plots using slits parallel to the spines of the RBEs. These plots, along with potential field extrapolation, led us to suggest that the occurrence of RBEs is generally correlated with the appearance of new, mixed or unipolar fields in close proximity to network fields. RBEs show a tendency to occur at the interface between large-scale fields and small-scale dynamic magnetic loops and thus are likely to be associated with existence of a magnetic canopy. Detection of kinked and/or inverse ``Y'' shaped RBEs further confirm this conclusion. 

\end{abstract}

\comments{Key words: Sun: atmosphere -- Sun: chromosphere -- Sun: oscillations}

\section{Introduction}

\noindent Type II spicules \citep{2007PASJ...59S.655D} observed in the chromosphere are thought to be, at least in part, responsible for the transfer of mass from the dense chromosphere into the corona and are suggested to play an important role in coronal heating and solar wind acceleration processes \citep[e.g.,][]{2011Sci...331...55D,bart_roots,scot_upflows}. Data further indicate that the type II spicules have a disk counterpart, so-called rapid blue-shifted excursions (RBEs), associated with the network fields \citep{langangen_2008, counterparts}. \ha\ RBEs are continuation of Ca II 854.2~nm RBEs, which tend to be located closer to the network fields \citep{Sekse_2012}. 

Type II spicules (hereafter spicules II) appear to possess physical properties different from classical spicules: a shorter lifetime (between 10~s and 100~s), smaller width (150 -- 700~km) as well as much higher line-of-sight (50 -- 150~km s$^{-1}$) and transverse (10~km s$^{-1}$) velocities \citep{Sekse_2013}. Spectroscopic studies also indicate that the plasma in these events is heated throughout their lifetime and the related upflows exhibit jet-like properties \citep{counterparts, Kuridze_2011}.

Spicules II are also known to exhibit rotational as well as transverse/swaying, volume filling motions \citep[e.g.,][]{Tomczyk_2007, Scott_2011Natur,torsional_osc} with amplitudes of 10-30~km s$^{-1}$ and periods of 100-500~s (see review by \cite{2009SSRv..149..355Z}). The displacements are perpendicular to spicules' axis and are interpreted as upward or downward propagating Alfv{\'e}n-like motions \citep{Okamoto_Bart}, or MHD kink mode waves \citep[e.g.,][]{2009ApJ...705L.217H, 0004-637X-749-1-30}.

There is little consensus among researchers on the origin of type II spicules. While some works suggest that reconnection process \citep{2008ApJ...679L..57I,2007PASJ...59S.655D,archontis_jet_model}, and oscillatory reconnection \citep{0004-637X-749-1-30} in particular, may account for their origin, others propose that  strong Lorentz force \citep{2011ApJ...736....9M} or propagation of the p-modes \citep{2009ApJ...702L.168D} may be the origin. Moreover, \cite{2011ApJ...730L...4J, Judge_2012} argues that spicules II could be warps in 2D sheet-like structures, while \cite{zhang12revision} questions the existence of spicules II as a distinct class altogether. The related difficulties in the interpretation of solar data mainly arise from the limited spatial resolution and complexity of the chromosphere \citep[e.g.,][]{2011ApJ...730L...4J,0004-637X-749-2-136}.

In this study we focus on the relationship between the occurrence of RBEs and the changes in the underlying dynamics of photospheric magnetic fields. We use \ha-0.1~nm and magnetic field data obtained with the New Solar Telescope \citep[NST,][]{goode_nst_2010, Cao_IRIM} installed at the Big Bear Solar Observatory (BBSO). 

\section{Observations, Data Reduction and Analysis}

\noindent On September 8, 2011 and August 14, 2011 the NST acquired quiet Sun data near the North Pole and in a coronal hole (CH) located near the solar equator at heliocentric-cartesian position (460"45"). In both cases the acquired data included blue wing \ha-0.1~nm images taken with a Zeiss Lyot filter with a 0.025~nm bandpass. The August 14, 2012 coronal hole data set spans nearly 100~min time interval with two 8~min data gaps (see Figs. 2 and 3) and it was complemented by magnetic field measurements performed with the InfraRed Imaging Magnetograph \cite[IRIM,][]{Cao_IRIM}.  \ha\ data were speckle reconstructed \citep{kisip_code}, aligned and de-stretched to remove residual image distortion due to seeing and telescope jitter. The intensity of each image was adjusted to the average level of the set and the final image pixel size was 0.052 arcsec.

IRIM is an imaging solar spectro-polarimeter that uses a pair of Zeeman sensitive Fe I lines present in the near infrared at 1564.85~nm and 1565.29~nm. The system is based on a 2.5~nm interference filter, a unique 0.025~nm birefringent Lyot filter, and a Fabry-P{\'e}rot etalon and it is capable of providing a bandpass as low as 0.01~nm over a field of view (FOV) of 50'' $\times$ 25''. With the aid of adaptive optics, we obtained circular and linear polarization images with spatial and temporal resolution of 0".25 and 43 s, respectively. In this study we focus on the morphology of the magnetic field with only Stokes V data being used. 

Polarization calibration, based on Muller matrix, was employed for IRIM data to minimize the cross-talk \citep{Cao_IRIM,Cao_2012_NIRIS}. The current calibration successfully corrects the cross-talk from M3 to the focal plane, and the matrix elements for correction of Q or U into V are of order of 10$^{-3}$. The correction of cross-talk introduced by the primary and secondary mirrors causes only a small error ($<$6-8\% of the analysed signal) and was not attempted. To convert IRIM data numbers into physical units (Gauss) we used co-temporal and co-spatial Hinode SOT/NFI measurements. The two data sets were carefully co-aligned and the best fit to the corresponding regression plot produced the coefficient of conversion.

\section{Results}

\begin{figure}[p]
\centering
\begin{tabular}{c}
\epsfxsize=6.0truein  \epsffile{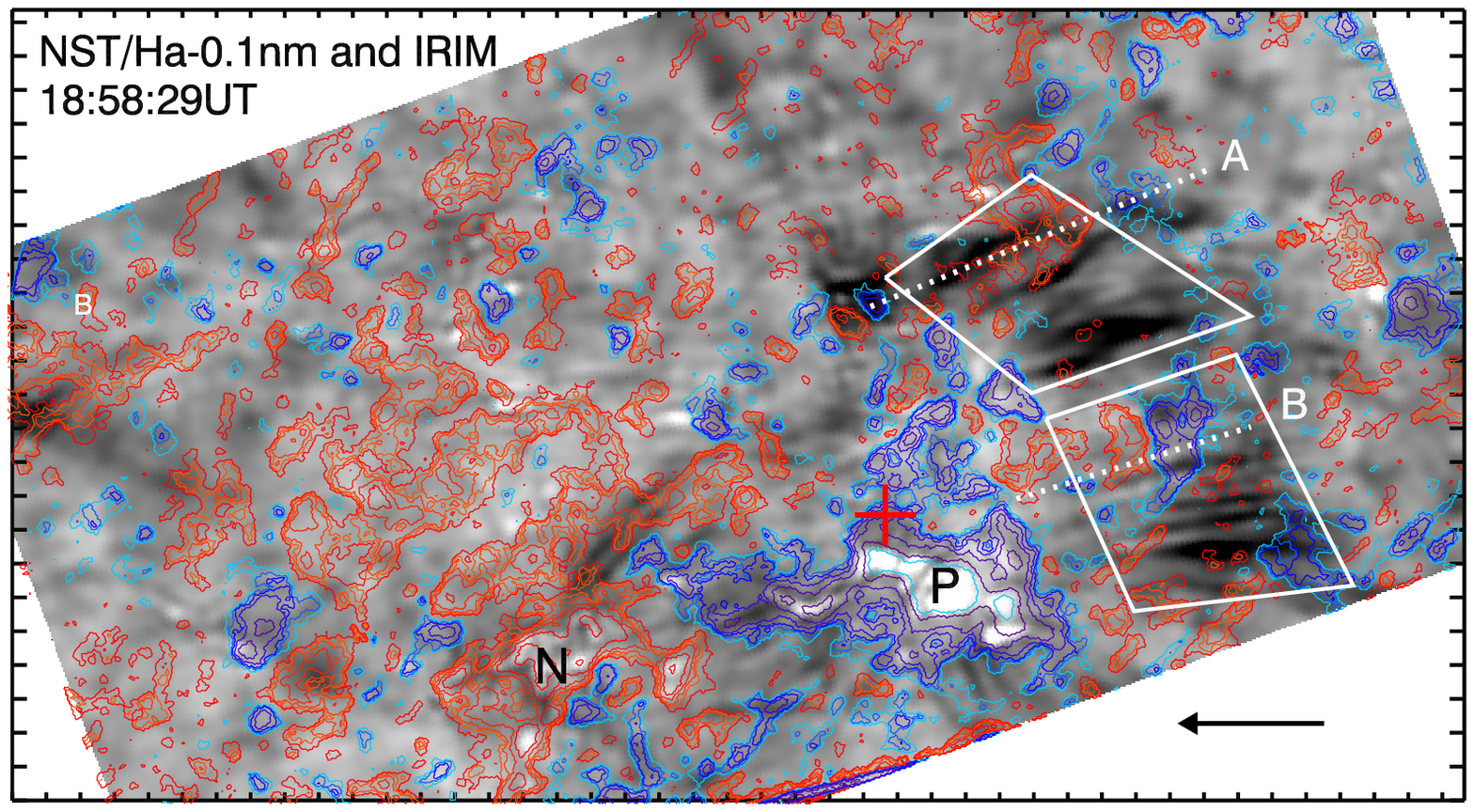} \
\end{tabular}
\caption{IRIM line-of-sight magnetic fields (contours) plotted over a co-temporal \ha-0.1~nm image. In this image the bright patches are clusters of magnetic bright points, while the dark jet-like features are rapid blue-shifted events (RBEs). The blue/red contours outline positive/negative polarity and are drawn at 20, 30, 40, 60, 100, 200, and 300~G levels. The two dotted lines mark the location of x-t plots shown in Figures \ref{cuta} and \ref{cutb}. The red cross indicates the location of a flux emergence episode discussed in Fig. \ref{case}. The location of the most persisting RBE activity is enclosed with boxes. The arrow in the lower right corner points toward the disk center. The tick marks separate 1~Mm intervals.}
\label{ha_irim}
\end{figure}

The August 14, 2011 data set is of undisturbed photosphere inside a CH.  Two small clusters of BPs, marked as N and P (Figure \ref{ha_irim}, top panel), were associated with opposite polarity fields and connected by a system of dynamic loops that were intermittently filled with jet-like plasma flows. The locations of the most persistent RBE activity (darker areas), were not evenly distributed over the FOV: RBEs most frequently stemmed from two distinct areas enclosed by boxes. The exception is the RBE activity associated with closed field lines connecting two N and P clusters, which will not be considered here. The potential field model showed that the right side of P (i.e., the area enclosed by boxes) was dominated by small-scale ($<$ 3~Mm) loops, while large-scale loop system was present on the left side of P. This asymmetry may lead to the formation of an extended magnetic canopy above the small-scale emerging dynamic fields on the right side of cluster P.

\begin{figure}[p]
\centering
\begin{tabular}{c}
\epsfxsize=6.5truein  \epsffile{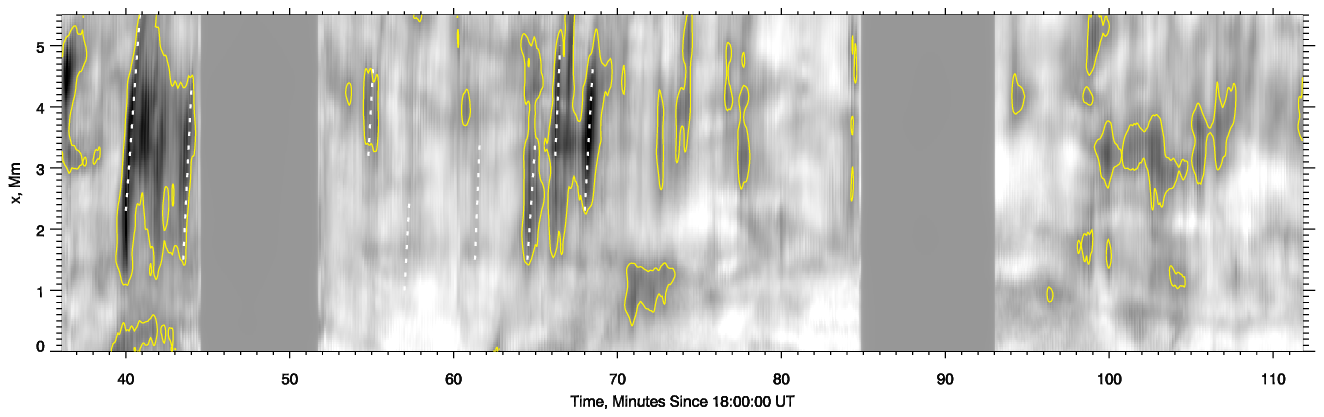} \\
\epsfxsize=6.5truein  \epsffile{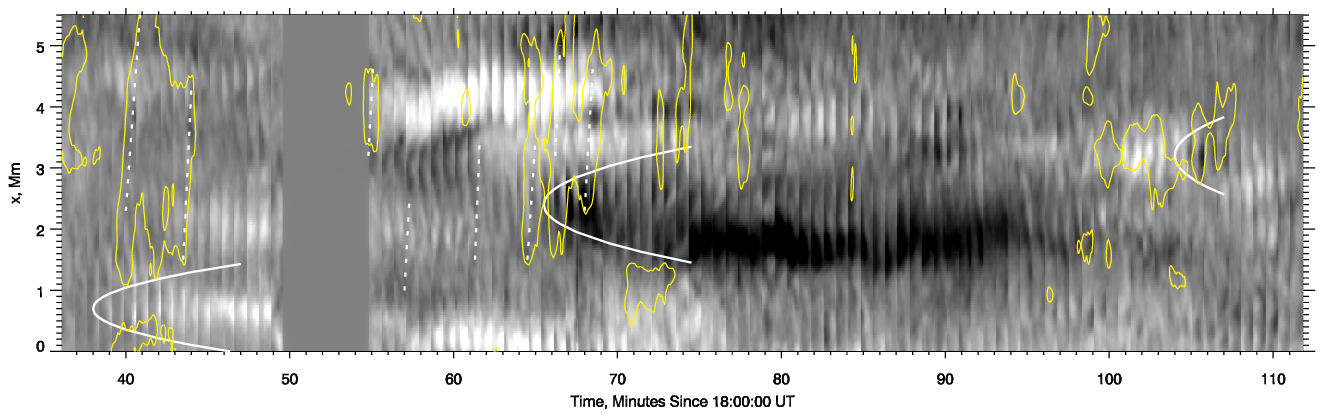}
\end{tabular}
\caption{An x-t plots made along slit A (Figure \ref{ha_irim}). The background in the top panel is an \ha-0.1~nm x-t plot and in the bottom panel is an IRIM Stokes V x-t plot. In total, 364 \ha-0.1~nm images were used to produce the top panel with the slit width being 6 pixels (0.23~Mm). The bottom panel was produced from 94 IRIM magnetograms and a 2.9~Mm wide slit was used. The darkest \ha-0.1~nm x-t features in the top panel were outlined at an arbitrary level by a yellow contour. The dashed line segments highlight steep x-t tracks representing the upward motion of RBEs. The lower end of the line segments (or dark tracks) marks their occurrence time and the initial position of the event along the slit. Similarly, the lower part of the encountered areas represent roots of RBEs, while the upper part are the RBE tips. The parabolas outline the several most prominent episodes of enhanced RBE activity associated with underlying dynamical changes in the magnetic field such as the appearance of mixed and/or unipolar fields.}
\label{cuta}
\end{figure}

\begin{figure}[p]
\centering
\begin{tabular}{c}
\epsfxsize=6.5truein  \epsffile{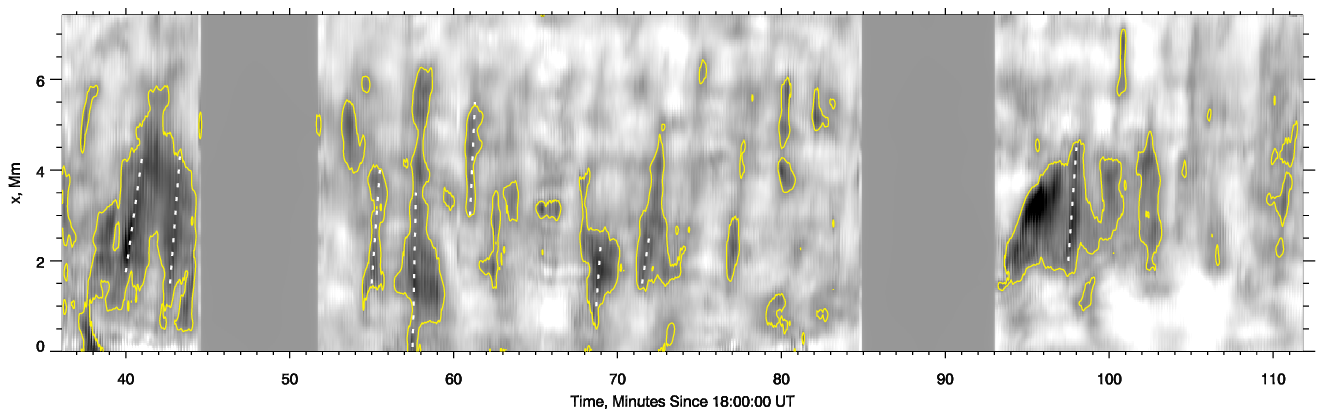} \\
\epsfxsize=6.5truein  \epsffile{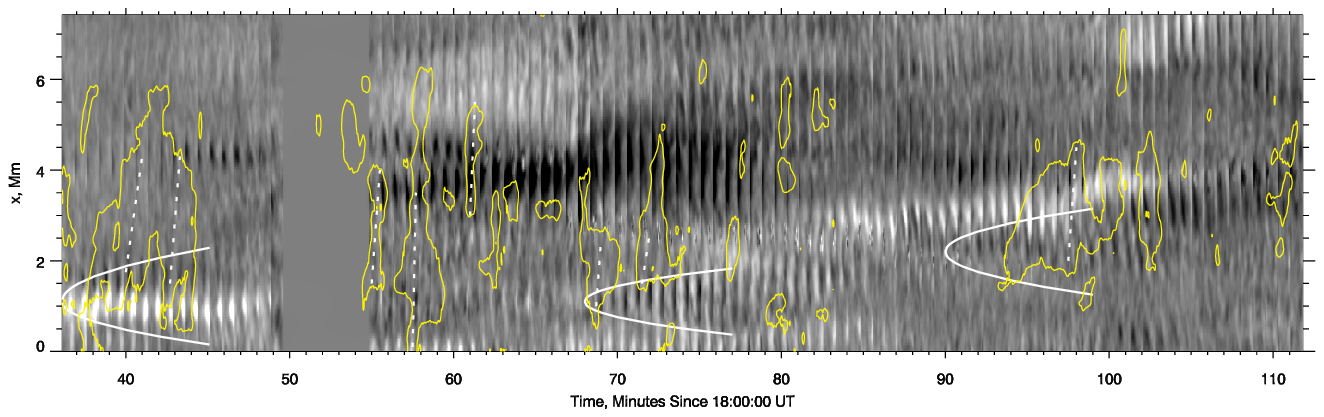}
\end{tabular}
\caption{x-t plots made for slit B (see Figure \ref{ha_irim}). The panel description is the same as in Figure \ref{cuta}.}
\label{cutb}
\end{figure}

In Figures \ref{cuta} and \ref{cutb}, we show x-t plots made along the cuts A and B (Figure \ref{ha_irim}, dotted lines). In each figure, the top panel is an \ha-0.1~nm x-t plot, while the bottom panel is an x-t plot of IRIM Stokes V data. A total of 364 \ha-0.1~nm images were used to produce the top panel with the slit width of 6 pixels (0.23~Mm). The bottom panel was produced from 94 IRIM magnetograms and a 2.9~Mm slit was used. In both top panels, the darkest \ha-0.1~nm x-t features were outlined by a yellow contour at an arbitrary intensity level. The dashed line segments highlight several steep x-t tracks representing an upward motion of RBEs. The lower end of the line segments (or dark tracks) marks the occurrence time and the initial position of the event along the slit. Similarly, the lower part of the areas, enclosed by the contours, represent roots of RBEs, while the upper part are RBE tips. We would like to stress that the level of RBE activity at these cuts is intermittent rather than constant: periods of enhanced RBE production are separated by long intervals of relative quietude (Figure \ref{cuta} and \ref{cutb}). A new period of RBE activity usually begins with the development of a dark jet-like feature, followed by series of increasingly thinner and fainter RBEs (see the \ha-0.1~nm x-t intervals between t=39~min and t=45~min, t=64~min and t=70~min in Figure \ref{cuta} and  t=38 -- 44~min and t=94 -- 99~min in Figure \ref{cutb}).

We overplotted the contours and tracks on the IRIM x-t plots shown in the lower panels of Figures \ref{cuta} and \ref{cutb}. There appears to be a tendency for the groups of RBEs  (and not necessary individual tracks) to occur when a new magnetic flux appears in the photosphere.  By using a parabola, we outlined the several most prominent episodes of enhanced RBE activity associated with underlying dynamical changes in the magnetic field such as the appearance of mixed and/or unipolar fields.

The first parabola in Figure \ref{cuta} has its apex set at x=0.7~Mm and t=38~min, when clear and persisting magnetic field changes began to occur at a location x. In this particular case, the magnetic field at x=0.7 Mm was mostly weak (grey) before  positive (white) polarity patch appeared at approx t=38 min. As the patch grew (t=40-44~min), some negative polarity fields appeared within the slice as well. At the same time, series of RBEs occurred with their long and dark, possibly highly inclined \ha-0.1~nm tracks stopping approx. 0.7~Mm short of the center of the positive polarity patch. Note that data, presented in \cite{Sekse_2012}, suggests that roots of Ca II 824.5~nm RBEs tend to be located lower in the chromosphere and closer to magnetic field concentrations, as compared to the roots of \ha\ RBEs. The difference can be as large as 1.5~Mm so that the real plasma outflows in the above example may originated much closer to the network flux concentrations.

The apex of the second parabola marks emergence of a negative polarity (black) fields. It is not clear whether two sets of RBEs that occurred between t=64 and t=67 min are related to this emergence episode since the t=66 min track could have originated at x=3.2 Mm as the dashed line suggests (top panel). However, the third group of RBEs that began at t=68~min seems to be temporally and spatially correlated with the new negative flux. Note that an emerging bipole may not always have its poles of equal intensity, or its axis parallel to the cut lines shown in Fig. 1. In this case only one polarity will appear in the slice, while the other polarity could be either outside the slice, masked by the surrounding the same polarity fields, or be much weaker and less pronounced. However, the most important observation is that the appearance of a new (bi- or mono-polar) flux at a given location, which suggests ongoing restructuring of the surrounding chromospheric fields as the new flux makes its way into the solar atmosphere.

The first parabola in Figure \ref{cutb} has its apex at t=36~min, when we assume the new bipole was beginning to emerge. The bipole showed signs of development (flux increase, some pole separation) and a prolonged series of RBEs appeared in association with the emergence. The second parabola in Figure  \ref{cutb} at t=67~min and x=1~Mm envelops what looks like an emerging bipole associated with two sets of enhanced RBE activity. The RBE activity associated with the third flux emergence event (x=2.5~Mm, t=87~min) was significantly (8~min) delayed relative to the first appearance of the negative polarity patch.

A common denominator in all these events is that the emergence of a new magnetic flux seems to be  accompanied by enhanced production of RBEs, whereas essentially no, or very weak, RBE activity is observed in ``gray'' areas occupied by the unipolar or weak mixed fields (e.g., between t=55~min and t=85~min and x $<$ 1.5~Mm, Figure \ref{cuta} and between t=62~min and t=75~min and x $>$ 5~Mm, Figure \ref{cutb}). These figures also suggest that RBE activity tends be delayed and often starts 2-5~min after the first signatures of the emerging flux appear. This time lag is long enough for emerging loops to reach the chromosphere, assuming that their average ascending rate is about 12~$km s^{-1}$ or higher \citep{martinez2009, martinez2010, Keys_2013}. We thus argue that RBE (spicule II) activity analysed here was caused by emergence of small-scale dipoles in close proximity to and around network field concentrations. 

\begin{figure}[p]
\centerline{\epsfxsize=5truein  \epsffile{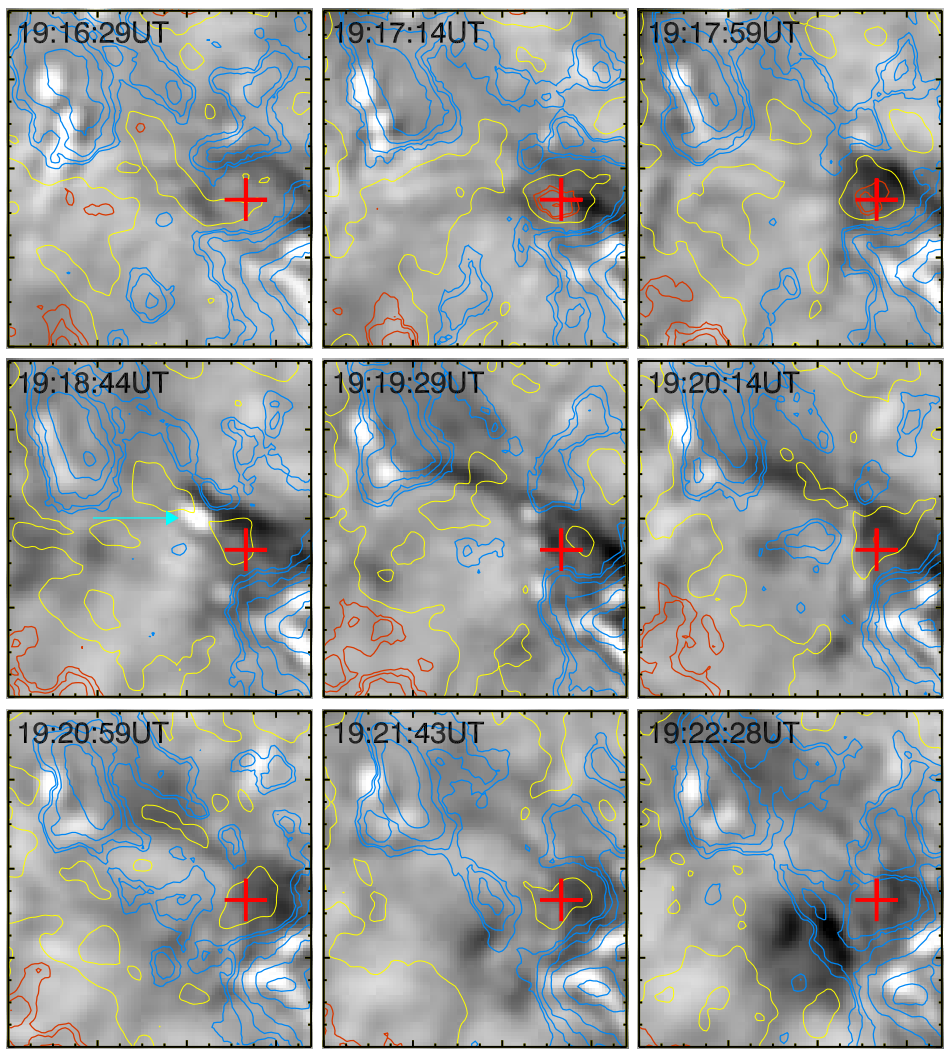}}
\caption{Background: A sequence of \ha-0.1~nm images. The dark jet-like feature in the 19:18:44~UT panel is an RBE. Blue/red contours represent IRIM line-of-sight magnetic fields plotted at $\pm$ 30, 40, 60, 100, 200, 300~G. The noise level in this particular data set was estimated to be 15~G. The yellow contour marks the polarity reversal line. The red cross marks emergence of a new negative flux concentration. The same location is also indicated in Fig. \ref{ha_irim}. The blue arrow points toward a transient brightening associated with the emergence and the jet. The solar limb is up. The large tick marks separate 1~Mm intervals.}
\label{case}
\end{figure}

In Figure \ref{case} we show an example of RBE activity (background) and the underlying evolution of magnetic fields (contours). The first contour is plotted at the level of $\pm$30~G, which corresponds to the 2$\sigma$ level of the data set. The red cross in Figure \ref{case} mark the location of emergence of a new negative polarity element, which occurred within the positive polarity fields at the edge of network cluster P  (also marked in Figure \ref{ha_irim}). The new element rapidly developed between 19:16:29~UT and 19:17:14~UT, and was later accompanied by a dark jet like feature that peaked around 19:18:44~UT and a short lived ($<$ 2~min) \ha-0.1~nm brightening (arrow). By that time the negative polarity element has already weakened, and it has disappeared by 19:22:28~UT.

\cite{2008ApJ...679L..57I} reported that the simulated emerging flux may be subject to magnetic reconnection with the pre-existing fields, which results in plasma acceleration and emission of high-frequency waves that propagate into the corona. Based on that agreement and the data presented above,  we argue that at least some RBEs (spicules II) are due to reconnection between the ubiquitous small-scale emerging fields \citep[e.g.,][]{centeno2007, lites_2008} and the pre-existing ``open'' (or large-scale closed) fields associated with network clusters.

\begin{figure}[p]
\centerline{\epsfxsize=5truein  \epsffile{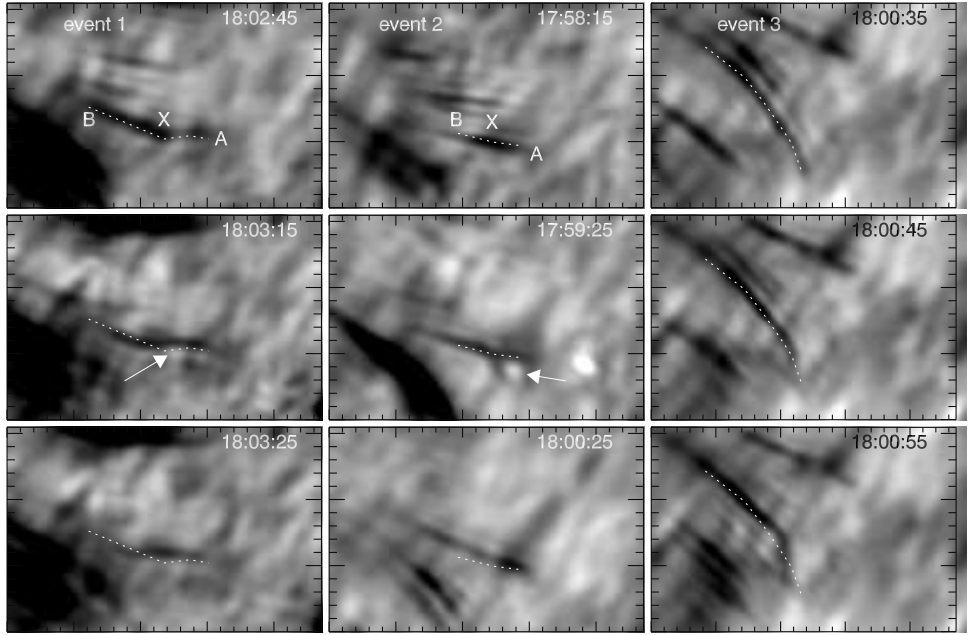}}
\caption{Evolution of three RBE events that exhibit a kink and an inverted ``Y'' shape, which are signatures of reconnection. These jets were rooted at point A, the point X may signify the field lines bending above the reconnection region, while the arc between A and X may be overlying the emerging bipole. Point B marks the dynamic top part of the jet event. The dotted line in each panel represents the shape of the feature as captured in the top panel. The large tick marks separate 1~Mm intervals. The solar limb is at the top.}
\label{kinks}
\end{figure}

In Figure \ref{kinks} we show three jet events that exhibit signatures of magnetic reconnection. These images are from the \ha-0.1~nm data set acquired near the solar North Pole. The shape of these jets is similar to those produced in simulations of reconnection of emerging flux and pre-existing large-scale fields (see e.g., Figure 3 in \cite{shibata2007} and Figure 1 in \cite{2008ApJ...679L..57I}). Event 1 \citep[see also][]{osc_paper} appears to be rooted at point A. Point X marks the location where field lines bend above what might be a reconnection point, and the arc between A and X is spanning the presumed emerging dipole. At about 18:03~UT a small low contrast brightening (arrow in the middle panel) developed just below X. Shortly after, point B retracted back and the jet faded from the FOV. Event 2 (middle column) appeared to be an inverted ``Y'' shaped jet, which is a signature of reconnection as well \citep{shibata2007}. It first appeared as a straight short RBE, while one minute later a brightening (arrow) and a small second leg have developed. It is unlikely that the second leg of an inverted ``Y'' was an intergranular lane, because of its short life time ($\sim$2~min) and synchronous evolution with the jet. Event 3 (right column) initially appeared as a typical curved RBE, however, it has later developed a kink (middle panel) that rapidly evolved. We would like to point out that because of their extremely short life time, these events were seen only in 3-4 consecutive frames taken 10~s apart. All events faded from the FOV some 20~s after the time of the last corresponding frames in Figure \ref{kinks}.

\section{Conclusions}

In this Letter, we present observational evidence that support the idea of the origin of RBEs (type II spicules) being in reconnection. The data suggest that occurrence of series of RBEs is generally correlated with instances of new flux emergence near the edge of a cluster of unipolar network fields. In this case, a large and dark jet-like features first appear, followed by numerous shorter and thinner jets. More specifically, we report that in these particular data sets i) RBEs tend to occur at an interface between large-scale (``open'') flux of a network fields and small-scale dynamic magnetic loops; ii) RBE occurrence is likely to be associated with the existence of a magnetic canopy; iii) there is a temporal correlation between the  appearance of new magnetic flux and periods on enhanced RBE activity, and iv) some RBEs display kink or inverse ``Y'' shape.

The data suggests that at least some of RBEs (type II spicules) are due to the interchange type of reconnection between the large-scale fields and the emerging small-scale fields. Namely, emerging magnetic dipoles reconnect with the ambient network fields in the manner of a standard X-ray jet or a blow-out jet \citep{Moore_RBE}. We further speculate that the reconnection may proceed in the form of oscillatory reconnection consisting of series of elementary reconnection events each generating plasma outflows, and high frequency waves \citep{0004-637X-749-1-30}. An additional possibility may be that reconnection with emerging dipoles removes flux from a bundle of flux tubes, thus inducing their rapid equilibrium reconfiguration. The disturbance may additionally generate both small scale (component) reconnection and high frequency MHD waves as proposed by \cite{0004-637X-736-1-3}, which may (along with oscillatory reconnection) account for the appearance of smaller and thinner spicules II as well as their group oscillations. 

Authors thank T.~J. Okamoto, B. de Pontieu, T.~J. Wang and H. Tian for valuable discussions. This work was conducted as part of the effort of NASA's Living with a Star Focused Science Team ``Jets''.  We thank BBSO observing and engineering staff for support and observations. This research was supported by NASA grants GI NNX08AJ20G, LWS NNX08AQ89G,  NNX11AO73G, as well as NSF AGS-1146896 grant.

\clearpage

\begin{thebibliography}{37}
\expandafter\ifx\csname natexlab\endcsname\relax\def\natexlab#1{#1}\fi

\bibitem[{{Archontis} {et~al.}(2010){Archontis}, {Tsinganos}, \&
  {Gontikakis}}]{archontis_jet_model}
{Archontis}, V., {Tsinganos}, K., \& {Gontikakis}, C. 2010, Astron. Astrophys.,
  512, L2

\bibitem[{{Cao} {et~al.}(2012){Cao}, {Goode}, {Ahn}, {Gorceix}, {Schmidt}, \&
  {Lin}}]{Cao_2012_NIRIS}
{Cao}, W., {Goode}, P.~R., {Ahn}, K., {Gorceix}, N., {Schmidt}, W., \& {Lin},
  H. 2012, in Astronomical Society of the Pacific Conference Series, Vol. 463,
  Astronomical Society of the Pacific Conference Series, ed. T.~R. {Rimmele},
  A.~{Tritschler}, F.~{W{\"o}ger}, M.~{Collados Vera}, H.~{Socas-Navarro},
  R.~{Schlichenmaier}, M.~{Carlsson}, T.~{Berger}, A.~{Cadavid}, P.~R.
  {Gilbert}, P.~R. {Goode}, \& M.~{Kn{\"o}lker}, 291

\bibitem[{{Cao} {et~al.}(2010){Cao}, {Gorceix}, {Coulter}, {Ahn}, {Rimmele}, \&
  {Goode}}]{Cao_IRIM}
{Cao}, W., {Gorceix}, N., {Coulter}, R., {Ahn}, K., {Rimmele}, T.~R., \&
  {Goode}, P.~R. 2010, Astronomische Nachrichten, 331, 636

\bibitem[{{Centeno} {et~al.}(2007){Centeno}, {Socas-Navarro}, {Lites}, {Kubo},
  {Frank}, {Shine}, {Tarbell}, {Title}, {Ichimoto}, {Tsuneta}, {Katsukawa},
  {Suematsu}, {Shimizu}, \& {Nagata}}]{centeno2007}
{Centeno}, R., {et~al.} 2007, \apjl, 666, L137

\bibitem[{{De Pontieu} {et~al.}(2012){De Pontieu}, {Carlsson}, {Rouppe van der
  Voort}, {Rutten}, {Hansteen}, \& {Watanabe}}]{torsional_osc}
{De Pontieu}, B., {Carlsson}, M., {Rouppe van der Voort}, L.~H.~M., {Rutten},
  R.~J., {Hansteen}, V.~H., \& {Watanabe}, H. 2012, \apjl, 752, L12

\bibitem[{{De Pontieu} {et~al.}(2009){De Pontieu}, {McIntosh}, {Hansteen}, \&
  {Schrijver}}]{bart_roots}
{De Pontieu}, B., {McIntosh}, S.~W., {Hansteen}, V.~H., \& {Schrijver}, C.~J.
  2009, \apjl, 701, L1

\bibitem[{{de Pontieu} {et~al.}(2007){de Pontieu}, {McIntosh}, {Hansteen},
  {Carlsson}, {Schrijver}, {Tarbell}, {Title}, {Shine}, {Suematsu}, {Tsuneta},
  {Katsukawa}, {Ichimoto}, {Shimizu}, \& {Nagata}}]{2007PASJ...59S.655D}
{de Pontieu}, B., {et~al.} 2007, \pasj, 59, 655

\bibitem[{{De Pontieu} {et~al.}(2011){De Pontieu}, {McIntosh}, {Carlsson},
  {Hansteen}, {Tarbell}, {Boerner}, {Martinez-Sykora}, {Schrijver}, \&
  {Title}}]{2011Sci...331...55D}
{De Pontieu}, B., {et~al.} 2011, Science, 331, 55

\bibitem[{{de Wijn} {et~al.}(2009){de Wijn}, {McIntosh}, \& {De
  Pontieu}}]{2009ApJ...702L.168D}
{de Wijn}, A.~G., {McIntosh}, S.~W., \& {De Pontieu}, B. 2009, \apjl, 702, L168

\bibitem[{{Goode} {et~al.}(2010){Goode}, {Coulter}, {Gorceix}, {Yurchyshyn}, \&
  {Cao}}]{goode_nst_2010}
{Goode}, P.~R., {Coulter}, R., {Gorceix}, N., {Yurchyshyn}, V., \& {Cao}, W.
  2010, Astronomische Nachrichten, 88, 789

\bibitem[{{He} {et~al.}(2009){He}, {Marsch}, {Tu}, \&
  {Tian}}]{2009ApJ...705L.217H}
{He}, J., {Marsch}, E., {Tu}, C., \& {Tian}, H. 2009, \apjl, 705, L217

\bibitem[{{Isobe} {et~al.}(2008){Isobe}, {Proctor}, \&
  {Weiss}}]{2008ApJ...679L..57I}
{Isobe}, H., {Proctor}, M.~R.~E., \& {Weiss}, N.~O. 2008, \apjl, 679, L57

\bibitem[{{Judge} {et~al.}(2012){Judge}, {Reardon}, \& {Cauzzi}}]{Judge_2012}
{Judge}, P.~G., {Reardon}, K., \& {Cauzzi}, G. 2012, \apjl, 755, L11

\bibitem[{{Judge} {et~al.}(2011){Judge}, {Tritschler}, \& {Chye
  Low}}]{2011ApJ...730L...4J}
{Judge}, P.~G., {Tritschler}, A., \& {Chye Low}, B. 2011, \apjl, 730, L4

\bibitem[{{Keys} {et~al.}(2013){Keys}, {Mathioudakis}, {Jess}, {Shelyag},
  {Christian}, \& {Keenan}}]{Keys_2013}
{Keys}, P.~H., {Mathioudakis}, M., {Jess}, D.~B., {Shelyag}, S., {Christian},
  D.~J., \& {Keenan}, F.~P. 2013, \mnras, 428, 3220

\bibitem[{{Kuridze} {et~al.}(2011){Kuridze}, {Mathioudakis}, {Jess}, {Shelyag},
  {Christian}, {Keenan}, \& {Balasubramaniam}}]{Kuridze_2011}
{Kuridze}, D., {Mathioudakis}, M., {Jess}, D.~B., {Shelyag}, S., {Christian},
  D.~J., {Keenan}, F.~P., \& {Balasubramaniam}, K.~S. 2011, \aap, 533, A76

\bibitem[{{Langangen} {et~al.}(2008){Langangen}, {De Pontieu}, {Carlsson},
  {Hansteen}, {Cauzzi}, \& {Reardon}}]{langangen_2008}
{Langangen}, {\O}., {De Pontieu}, B., {Carlsson}, M., {Hansteen}, V.~H.,
  {Cauzzi}, G., \& {Reardon}, K. 2008, \apjl, 679, L167

\bibitem[{Leenaarts {et~al.}(2012)Leenaarts, Carlsson, \& van~der
  Voort}]{0004-637X-749-2-136}
Leenaarts, J., Carlsson, M., \& van~der Voort, L.~R. 2012, The Astrophysical
  Journal, 749, 136

\bibitem[{{Lites} {et~al.}(2008){Lites}, {Kubo}, {Socas-Navarro}, {Berger},
  {Frank}, {Shine}, {Tarbell}, {Title}, {Ichimoto}, {Katsukawa}, {Tsuneta},
  {Suematsu}, {Shimizu}, \& {Nagata}}]{lites_2008}
{Lites}, B.~W., {et~al.} 2008, \apj, 672, 1237

\bibitem[{{Mart{\'i}nez Gonzalez} \& {Bellot Rubio}(2009)}]{martinez2009}
{Mart{\'i}nez Gonzalez}, M.~J., \& {Bellot Rubio}, L.~R. 2009, \apj, 700, 1391

\bibitem[{{Mart{\'i}nez Gonzalez} {et~al.}(2010){Mart{\'i}nez Gonzalez}, {Manso
  Sainz}, {Asensio Ramos}, \& {Bellot Rubio}}]{martinez2010}
{Mart{\'i}nez Gonzalez}, M.~J., {Manso Sainz}, R., {Asensio Ramos}, A., \&
  {Bellot Rubio}, L.~R. 2010, \apjl, 714, L94

\bibitem[{{Mart{\'{\i}}nez-Sykora} {et~al.}(2011){Mart{\'{\i}}nez-Sykora},
  {Hansteen}, \& {Moreno-Insertis}}]{2011ApJ...736....9M}
{Mart{\'{\i}}nez-Sykora}, J., {Hansteen}, V., \& {Moreno-Insertis}, F. 2011,
  \apj, 736, 9

\bibitem[{{McIntosh} \& {De Pontieu}(2009)}]{scot_upflows}
{McIntosh}, S.~W., \& {De Pontieu}, B. 2009, \apj, 707, 524

\bibitem[{{McIntosh} {et~al.}(2011){McIntosh}, {de Pontieu}, {Carlsson},
  {Hansteen}, {Boerner}, \& {Goossens}}]{Scott_2011Natur}
{McIntosh}, S.~W., {de Pontieu}, B., {Carlsson}, M., {Hansteen}, V., {Boerner},
  P., \& {Goossens}, M. 2011, \nat, 475, 477

\bibitem[{{McLaughlin} {et~al.}(2012){McLaughlin}, {Verth}, {Fedun}, \&
  {Erd{\'e}lyi}}]{0004-637X-749-1-30}
{McLaughlin}, J.~A., {Verth}, G., {Fedun}, V., \& {Erd{\'e}lyi}, R. 2012, The
  Astrophysical Journal, 749, 30

\bibitem[{{Moore} {et~al.}(2011){Moore}, {Sterling}, {Cirtain}, \&
  {Falconer}}]{Moore_RBE}
{Moore}, R.~L., {Sterling}, A.~C., {Cirtain}, J.~W., \& {Falconer}, D.~A. 2011,
  \apjl, 731, L18

\bibitem[{{Okamoto} \& {De Pontieu}(2011)}]{Okamoto_Bart}
{Okamoto}, T.~J., \& {De Pontieu}, B. 2011, \apjl, 736, L24

\bibitem[{{Rouppe van der Voort} {et~al.}(2009){Rouppe van der Voort},
  {Leenaarts}, {de Pontieu}, {Carlsson}, \& {Vissers}}]{counterparts}
{Rouppe van der Voort}, L., {Leenaarts}, J., {de Pontieu}, B., {Carlsson}, M.,
  \& {Vissers}, G. 2009, \apj, 705, 272

\bibitem[{{Sekse} {et~al.}(2013){Sekse}, {Rouppe van der Voort}, \& {De
  Pontieu}}]{Sekse_2013}
{Sekse}, D.~H., {Rouppe van der Voort}, L., \& {De Pontieu}, B. 2013, \apj,
  764, 164

\bibitem[{Sekse {et~al.}(2012)Sekse, van~der Voort, \& Pontieu}]{Sekse_2012}
Sekse, D.~H., van~der Voort, L.~R., \& Pontieu, B.~D. 2012, The Astrophysical
  Journal, 752, 108

\bibitem[{{Shibata} {et~al.}(2007){Shibata}, {Nakamura}, {Matsumoto}, {Otsuji},
  {Okamoto}, {Nishizuka}, {Kawate}, {Watanabe}, {Nagata}, {UeNo}, {Kitai},
  {Nozawa}, {Tsuneta}, {Suematsu}, {Ichimoto}, {Shimizu}, {Katsukawa},
  {Tarbell}, {Berger}, {Lites}, {Shine}, \& {Title}}]{shibata2007}
{Shibata}, K., {et~al.} 2007, Science, 318, 1591

\bibitem[{{Tomczyk} {et~al.}(2007){Tomczyk}, {McIntosh}, {Keil}, {Judge},
  {Schad}, {Seeley}, \& {Edmondson}}]{Tomczyk_2007}
{Tomczyk}, S., {McIntosh}, S.~W., {Keil}, S.~L., {Judge}, P.~G., {Schad}, T.,
  {Seeley}, D.~H., \& {Edmondson}, J. 2007, Science, 317, 1192

\bibitem[{van Ballegooijen {et~al.}(2011)van Ballegooijen, Asgari-Targhi,
  Cranmer, \& DeLuca}]{0004-637X-736-1-3}
van Ballegooijen, A.~A., Asgari-Targhi, M., Cranmer, S.~R., \& DeLuca, E.~E.
  2011, The Astrophysical Journal, 736, 3

\bibitem[{{W{\"o}ger} \& {von der L{\"u}he}(2007)}]{kisip_code}
{W{\"o}ger}, F., \& {von der L{\"u}he}, O. 2007, Appl. Opt., 46, 8015

\bibitem[{{Yurchyshyn} {et~al.}(2012){Yurchyshyn}, {Kilcik}, \&
  {Abramenko}}]{osc_paper}
{Yurchyshyn}, V., {Kilcik}, A., \& {Abramenko}, V. 2012, ArXiv e-prints

\bibitem[{{Zaqarashvili} \& {Erd{\'e}lyi}(2009)}]{2009SSRv..149..355Z}
{Zaqarashvili}, T.~V., \& {Erd{\'e}lyi}, R. 2009, \ssr, 149, 355

\bibitem[{{Zhang} {et~al.}(2012){Zhang}, {Shibata}, {Wang}, {Mao}, {Matsumoto},
  {Liu}, \& {Su}}]{zhang12revision}
{Zhang}, Y.~Z., {Shibata}, K., {Wang}, J.~X., {Mao}, X.~J., {Matsumoto}, T.,
  {Liu}, Y., \& {Su}, J.~T. 2012, ArXiv e-prints

\end{thebibliography}
\end{document}